\documentstyle[preprint,aps]{revtex}
\begin{document}
\draft
\preprint{ANL-HEP-PR-95-88}
\title{Breakdown of Conventional Factorization \\
       for Isolated Photon Cross Sections }
\author{Edmond L. Berger$^1$, Xiaofeng Guo$^2$, and Jianwei Qiu$^2$}
\address{$^1$High Energy Physics Division,
             Argonne National Laboratory \\
             Argonne, Illinois 60439, USA \\
         $^2$Department of Physics and Astronomy,
             Iowa State University \\
             Ames, Iowa 50011, USA }
\date{December 11, 1995}
\maketitle

\begin{abstract}
Using $e^+e^-\rightarrow\gamma + X$ as an example, we show that the
conventional factorization theorem in perturbative quantum
chromodynamics breaks down for isolated photon cross sections in a
well defined part of phase space.  Implications and physical
consequences are discussed.
\end{abstract}
\vspace{0.2in}
\narrowtext


High energy photons have long been considered an excellent probe of
\underline{short-distance} physics in strong interactions.  They
couple directly to pointlike quark constituents and do not interact
much once produced.\cite{Owens}  Photons can also result from
\underline{long-distance} fragmentation of quarks and
gluons, themselves produced in short-distance hard collisions.
Consequently, the \underline{inclusive} photon cross section at
high energy includes both short-distance direct and long-distance
fragmentation contributions, and the cross section is not completely
perturbative.  Nevertheless, in accord with the factorization theorem of
perturbative quantum chromodynamics (QCD)\cite{CSS}, all long-distance
physics associated with parton-to-photon fragmentation can be
represented by non-perturbative, but well-defined and universal photon
fragmentation functions, and the remainder of the theoretical
expression for the cross section, calculable in QCD perturbation
theory, is insensitive to the infrared region of the theory.

However, for observational reasons the \underline{inclusive} cross
section may not be measurable at high energy. Owing to backgrounds
from, e.g., $\pi^0\rightarrow \gamma\gamma$, a single high energy
photon is observed and the cross section is measured only when
the photon is relatively isolated.  In the experimental definition of
isolation, a cone of half-angle $\delta$ is drawn about the direction
of the photon's momentum, and the cross section is measured for photons
accompanied by less than a specified amount of hadronic energy in the
cone, {\it e.g.}, $E_h^{cone}\leq E_{\rm max}$.  Because of isolation,
the experimental cross section for isolated photons depends explicitly
on the isolation parameters $\delta$ and $E_{\rm max}$.

A proper theoretical treatment of the cross section for
isolated photons requires careful consideration of the origins and
cancellation of both
infrared and collinear singularities in QCD perturbation theory.
In a theoretical calculation, isolation of the photon restricts the
final-state phase space accessible to accompanying quarks and gluons.
In this Letter, using $e^+e^-\rightarrow\gamma X$ as an
example, we demonstrate that this phase space restriction inevitably
breaks the perfect cancellation of infrared singularities between real
gluon emission and virtual gluon exchange diagrams that is required to
yield finite cross sections in each perturbative order.

Breakdown of the cancellation of infrared singularities appears first
at next-to-leading order in the fragmentation contributions.  The
associated physics can be summarized as follows.
In the fragmentation contribution, sketched in Fig.~1,
hadronic energy in the isolation cone has two sources: a)
energy from parton fragmentation, $E_{frag}$, and b) energy from
non-fragmenting final-state partons, $E_{partons}^{cone}$, that enter
the cone.  When the maximum hadronic energy allowed in the
isolation cone is saturated by the fragmentation energy, $E_{\rm
max}=E_{frag}$, there is no allowance for energy in the cone from other
final-state partons.  In particular, if there is a gluon in the final
state, the phase space for this gluon becomes restricted.  By
contrast, isolation does not affect the virtual
gluon exchange contribution.  Therefore, in the isolated case, there is
a possibility that the infrared singularity from the virtual contribution
may not be cancelled completely by the \underline{restricted} real
contribution.  In the
remainder of this Letter, we show that this is indeed the case, and we
explore the implications.

The cross section for the inclusive yield of high energy photons in
hadronic final states of $e^+e^-$ annihilation is well-defined, in
accord with the factorization theorem of perturbative QCD.\cite{BGQ1}
For isolated photons, we first assume that the factorization theorem
holds; we then follow standard procedures to calculate the
short-distance partonic hard parts perturbatively. Finally, we
demonstrate that some of the hard parts have uncanceled infrared
singularities.

If conventional factorization were true, the cross section for isolated
photons would be expressed in the following factorized form
\begin{eqnarray}
E_\gamma \frac{d\sigma^{iso}_{e^+e^- \rightarrow \gamma X}}{d^3\ell}
&=& \sum_c\
    \int_{{\rm max}\left[x_\gamma,\frac{1}{1+\epsilon_h}\right]}^1
         \frac{dz}{z}\
    E_c \frac{d\hat{\sigma}^{iso}_{e^+e^- \rightarrow cX}}{d^3 p_c}
    \left(x_c=\frac{x_\gamma}{z}\right)
    \frac{D_{c \rightarrow \gamma}(z,\delta)}{z}
    \nonumber \\
&\equiv & \sum_c\
    E_c \frac{d\hat{\sigma}^{iso}_{e^+e^- \rightarrow cX}}{d^3 p_c}
    \otimes D_{c \rightarrow \gamma}(z,\delta)\ .
\label{e1}
\end{eqnarray}
In Eq.~(\ref{e1}), $x_\gamma=2E_\gamma/\sqrt{s}$, $x_c=2E_c/\sqrt{s}$,
$\epsilon_h=E_{\rm max}/E_\gamma$, and the sum
extends over $c = \gamma,q,\bar{q}$ and $g$;
$D_{c\rightarrow\gamma}(z,\delta)$ is the
nonperturbative function that describes fragmentation of parton
``$c$'' into a photon.  Fragmentation is assumed theoretically to be a
collinear process.
The lower limit of the $z$-integration results from the
isolation requirement with the assumption that all fragmentation energy
is in the isolation cone.\cite{BGQ2}
For the short-distance partonic hard parts,
$E_c d\hat{\sigma}^{iso}_{e^+e^- \rightarrow cX}/d^3p_c$,
the isolation requirement is that the energy carried into the cone by
non-fragmenting partons (i.e., partons other than $c$) should satisfy
$E_{partons}^{cone} \leq E_{\rm min}(z)\equiv
[(1+\epsilon_h)-1/z]E_\gamma$.
To demonstrate the incomplete cancellation of infrared
singularities, we present the one-loop hard part for the quark
fragmentation contribution: $e^+e^-\rightarrow q X; q \rightarrow
\gamma$.  The derivation of all other contributions is found in
Ref.~\cite{BGQ2}.

To calculate the one-loop quark fragmentation contribution, we
consider $e^+e^-\rightarrow q X$ and apply Eq.~(\ref{e1})
perturbatively to first order in $\alpha_s$.  We derive
\begin{equation}
\hat{\sigma}^{(1)iso}_{e^+e^-\rightarrow q X}(x_q)
= \left. \sigma^{(1)iso}_{e^+e^-\rightarrow q X}
  \right|_{E_g(or\, E_{\bar{q}})\leq E_{\rm min}}
- \hat{\sigma}^{(0)iso}_{e^+e^-\rightarrow q' X}(x_q')
  {\otimes}\, D^{(1)}_{q'\rightarrow q}(z') \ .
\label{e3}
\end{equation}
The convolution over $z'$ in Eq.~(\ref{e3}) is the same as that
in Eq.~(\ref{e1}) but with $z$ replaced by $z'$, and the lower limit
replaced by\ $max[x_q,x_q/(x_\gamma(1+\epsilon_h))]$.  In
Eq.~(\ref{e3}), $\hat{\sigma}^{(0)iso}_{e^+e^-\rightarrow q' X}(x_q')$
is the zeroth order hard part obtained from the lowest order
Feynman diagram for $e^+e^-\rightarrow q\bar{q}$, and it is
finite;\cite{BGQ2}  $D^{(1)}_{q'\rightarrow q}(z)$
is the singular first order quark-to-quark fragmentation
function.\cite{BGQ1} On the right side of Eq.~(\ref{e3}), the first
order isolated partonic cross
section, $\sigma^{(1)iso}_{e^+e^-\rightarrow q X}$, has both
real gluon emission and virtual gluon exchange contributions.
The real contributions are obtained from three body final-state tree
diagrams $e^+e^-\rightarrow q\bar{q}g$, and the virtual contributions
from one-loop interference diagrams $e^+e^-\rightarrow q\bar{q}$.
Integration over the phase space of the gluon from $e^+e^-\rightarrow
q\bar{q}g$ yields both
infrared (when $E_g\rightarrow 0$) and collinear (when $g$ is
parallel to the observed $q$) divergences.  If factorization is
valid, the infrared divergence will be cancelled by the infrared
divergence from the virtual diagrams, and the collinear divergence
will be cancelled by the singular second term on the right side of
Eq.~(\ref{e3}).

Our calculation of $\sigma^{(1)iso}_{e^+e^-\rightarrow q X}$ proceeds
as follows.  For the real gluon emission (``$R$'') contribution we
write\cite{BGQ2}
\begin{equation}
d\sigma^{(R)iso} = \sum_{q} \left[\frac{2}{s}F^{PC}_q(s)\right]\
e^2\ C\ \frac{1}{4}H\ dPS^{(3)iso} .
\label{e7}
\end{equation}
The constant $C$ is an overall color factor, $e$ is the electric charge,
and the normalization factor
$(2/s)F_q^{PC}(s)$ includes contributions from both $\gamma^*$ and
$Z^0$ intermediate states.  We use dimensional regularization
with $n=4-2\epsilon$.  Letting $p_1$, $p_2$ and $p_3$ label the
momenta of the $q$, $\bar{q}$ and $g$, respectively, we derive the
squared matrix element $H$.
\begin{eqnarray}
\frac{1}{4}H
&=&
 \left( 1 + \cos^2 \theta_1 - 2\epsilon\right)
  \left[ \left( \frac{1+x_1^2}{1-x_1} \right) \frac{1}{y_{13}}
  + \frac{y_{13}}{1-x_1} \right. \nonumber \\
&&{\hskip 1.2in}
  \left. -\frac{2}{1-x_1}
         -\epsilon\left( \frac{1-x_1}{y_{13}}+
                         \frac{y_{13}}{1-x_1}+2 \right) \right]
         \nonumber \\
&+& \left( 1-3\cos^2\theta_1\right)
  \left[ \frac{2}{x_1} \left(1-\frac{y_{13}}{x_1}\right) \right] \ ,
\label{e8}
\end{eqnarray}
where $x_1=2E_1/\sqrt{s}$ ($=x_q$ in Eq.~(\ref{e3})),
$y_{13}=2p_1\cdot p_3/s$, and $\theta_1$ is the angle between the
``observed'' quark and the $e^+e^-$ beam axis.
An overall coupling constant $(e\mu^{\epsilon})^2 (g\mu^{\epsilon})^2$
is omitted in Eq.~(\ref{e8}).  The three particle phase space element is
\widetext
\begin{eqnarray}
dPS^{(3)iso} &=&
  \frac{1}{2}\frac{1}{(2\pi)^3}\frac{d^3p_1}{E_1}
  \left( {{4\pi} \over {(s/4) \sin^2\theta_1}}\right)^\epsilon
  {{1} \over {\Gamma(1-\epsilon)}}
  \frac{2\pi}{s}{{\delta\left( x_1-(1-y_{23})\right)} \over {x_1}}
  \label{e9}   \\
&\times &\frac{s}{4}\left[
  \left(\frac{1}{2\pi}\right)^2\left(\frac{4\pi}{s}\right)^\epsilon
  \frac{1}{\Gamma(1-\epsilon)} \right]
  \delta\left( 1-y_{12}-y_{13}-y_{23}\right)
  \frac{dy_{12}}{y^\epsilon_{12}}\,
  \frac{dy_{13}}{y^\epsilon_{13}}\,
  \frac{dy_{23}}{y^\epsilon_{23}}\,
  S(y_{12},y_{13},y_{23}) \ , \nonumber
\end{eqnarray}
\narrowtext
\noindent where $y_{ij}=2p_i\cdot p_j/s$ with $i,j=1,2,3$.
Function $S(y_{12},y_{13},y_{23})$ in Eq.~(\ref{e9}) specifies the
constraints on the phase space due to isolation.  The
calculation of the \underline{isolated} partonic cross section is
identical to that for the inclusive cross section, except for the
phase space integration.  Once the $\delta$-functions are used
one integral is left, say $dy_{13}$.  Isolation
places constraints on the limits of the $y_{13}$ integration.

In the inclusive case, the limits of integration over
$\hat{y}_{13}\equiv y_{13}/x_1$ are from 0 to 1.
Integration of the matrix element in Eq.~(\ref{e8})
over $\hat{y}_{13}$ from 0 to 1 generates poles in $1/\epsilon$.  As
examples, a $1/\epsilon^2$ singularity arises from the
$1/[(1-x_1)\hat{y}_{13}]$ term, and a $1/\epsilon$ singularity from
the $1/(1-x_1)$ and $1/\hat{y}_{13}$ terms.
After contributions are combined from the real and the virtual gluon
diagrams, all singularities cancel in the inclusive case, except one
$1/\epsilon$ term due to a collinear singularity between the
fragmenting quark and the real
gluon.\cite{BGQ1}  It is cancelled by the second term
on the right side of Eq.~(\ref{e3}).

In the isolated photon situation, isolation splits the integration over
$\hat{y}_{13}$ into three regions:
\begin{equation}
\int d\hat{y}_{13}\,
= \int_{0}^{\min[\bar{y}_c,\bar{y}_m]} d\hat{y}_{13}
+ \int_{\bar{y}_c}^{1-\bar{y}_c} d\hat{y}_{13}
+ \int_{\max[(1-\bar{y}_c),(1-\bar{y}_m)]}^{1} d\hat{y}_{13}\ .
\label{e10}
\end{equation}
Quantities $\bar{y}_c$ and $\bar{y}_m$ are
\begin{eqnarray}
\bar{y}_c &\equiv &
 \frac{(1-x_1)\sin^2(\delta/2)}{1-x_1\sin^2(\delta/2)}\ \
\Rightarrow\ \ (1-x_1)\frac{\delta^2}{4}\ ;  \nonumber \\
\bar{y}_m &\equiv &
 z(1+\epsilon_h)-\frac{1}{x_1}\
=\ z\left[ (1+\epsilon_h)-\frac{1}{x_\gamma}\right]\ ,
\label{e11}
\end{eqnarray}
with $z=x_\gamma/x_1$.  For the first interval in Eq.~(\ref{e10}),
the condition $0\leq \hat{y}_{13} \leq \bar{y}_c$ ensures that a gluon
is in the isolation cone of the fragmenting quark; and
$\hat{y}_{13}\leq \min[\bar{y}_c,\bar{y}_m]$ ensures that the total
hadronic energy in the isolation cone is less than $E_{\rm max}=
\epsilon_h E_\gamma$. Similarly, the condition $\max[(1-\bar{y}_c),
(1-\bar{y}_m)] \leq \hat{y}_{13} \leq 1$ for the third interval
ensures that the antiquark is in the isolation cone of the fragmenting
quark, and that the total hadronic
energy in the isolation cone is less than $E_{\rm max}$.  The second
interval represents the situation when neither gluon nor antiquark is
in the isolation cone.

Equations~({\ref{e10}) and (\ref{e11}) show that the isolated cross
section is identical to the inclusive cross section if $\delta=0$ or
$\bar{y}_c \leq \bar{y}_m$.  However, if $\bar{y}_c > \bar{y}_m$, the
phase space of the final state gluon (and/or antiquark) is
\underline{smaller}.

We now examine in turn two separate situations: $x_\gamma
\leq 1/(1+\epsilon_h)$ and $x_\gamma > 1/(1+\epsilon_h)$.  Both yield
isolated photon partonic cross sections that are infrared sensitive.
When $x_\gamma \leq 1/(1+\epsilon_h)$, $\bar{y}_m  \leq 0$, and only
the second interval in Eq.~(\ref{e10}) survives.  We reexpress it as
\begin{equation}
\int_{(1-x_1)\delta^2/4}^{1-(1-x_1)\delta^2/4} d\hat{y}_{13}
= \int_0^1 d\hat{y}_{13}
- \int_0^{(1-x_1)\delta^2/4} d\hat{y}_{13}
- \int_{1-(1-x_1)\delta^2/4}^1 d\hat{y}_{13}\ ,
\label{e12}
\end{equation}
where $\bar{y}_c$ is expanded to order $\delta^2$.
The \underline{first} term on the right side of Eq.~(\ref{e12}) is,
by definition, the complete real contribution to
the inclusive partonic cross section.  When it is combined with the
virtual contribution, as in the inclusive case, all pole terms
cancel, except for one $1/\epsilon$ term due to the collinear
singularity, discussed above.
The \underline{second} and \underline{third} terms diminish
the real contribution in isolated case.  The
third term is innocuous in that it does not generate a
$1/\epsilon$ singularity; it yields only terms which vanish as
$\delta^2\rightarrow 0$.  However, the second term
generates a number of pole terms in $1/\epsilon$.

Neglecting all terms of $O(\delta^2)$ or higher, we obtain
the isolated partonic cross section after including real gluon
contributions from all three terms in Eq.~(\ref{e12}) and the
virtual gluon contribution.
\newpage
\begin{eqnarray}
E_1 \frac{d\sigma^{(1)iso}_{e^+e^-\rightarrow q X}}{d^3p_1}
&=&
 \left[ \frac{2}{s}\, F^{PC} (s)\right]
 \left[ \alpha^2_{em}\, N_c\, \frac{1}{x_1} \right]
         C_F \left( \frac{\alpha_s}{2\pi} \right)
          \nonumber \\
&&\times \Bigg\{
  \left(1+\cos^2\theta_1 \right)
  \Bigg[-\left(1+x_1^2\right)\left(\frac{\ell n(1-x_1)}{1-x_1}\right)_+
   \nonumber \\
&&\quad\quad\quad
  -\frac{1+x_1^2}{(1-x_1)_+} \ell n\left(\frac{\delta^2}{4}\right)
  -\frac{3}{2}\frac{1}{(1-x_1)_+}
  -\frac{1}{2}\left(x_1-3\right)
   \nonumber \\
&&\quad\quad\quad
    +\, \delta(1-x_1)\left[\frac{2\pi^2}{3}-\frac{9}{2}
       -\frac{1}{2}\ell n^2\left(\frac{\delta^2}{4}\right)\right]
   \Bigg]
   \nonumber \\
&&\quad + \left(1-3\cos^2\theta_1\right) \Bigg\}
   \nonumber \\
&-& \left[ \frac{2}{s}\, F^{PC} (s)\right]
    \left[ \alpha^2_{em}\, N_c
    \left( \frac{4\pi \mu^2}{(s/4) \sin^2\theta_1}\right)^\epsilon
           \frac{1}{\Gamma(1-\epsilon)}\, \frac{1}{x_1}\, \right]
   \nonumber \\
&&\times
    \left( 1 + \cos^2 \theta_1 - 2\epsilon \right)\,
    C_F \left( \frac{\alpha_s}{2\pi}\right)
    \left[ \left(\frac{4\pi \mu^2}{s}\right)^\epsilon
           \frac{1}{\Gamma(1-\epsilon)} \right]
   \nonumber \\
&&\times
    \left\{\frac{1}{\epsilon^2}+\frac{1}{\epsilon}
          \left(\frac{3}{2}-\ell n\frac{\delta^2}{4}\right)
         \right\}\, \delta(1-x_1)\ .
\label{e15}
\end{eqnarray}
As explained above, the uncanceled poles in Eq.~(\ref{e15}) come from
the interval specified by the \underline{second} term in
Eq.~(\ref{e12}).  The singularities corresponding to these poles are
infrared in nature and, as expected, are proportional to
$\delta(1-x_1)$.  These poles would be irrelevant if $x_1\neq 1$.
However, $x_1\equiv x_\gamma/z$, and $x_1=1$ is kinematically allowed
when $x_\gamma=1/(1+\epsilon_h)$.
Since the subtraction term in Eq.~(\ref{e3}) vanishes when $x_\gamma
\leq 1/(1+\epsilon_h)$, the poles in Eq.~(\ref{e15}) become the
uncanceled infrared singularities of the short-distance hard part,
$\hat{\sigma}_{e^+e^-\rightarrow q X}^{(1)iso}$ defined in
Eq.~(\ref{e3}).

Having examined the situation for $x_\gamma
\leq 1/(1+\epsilon_h)$, we next consider $x_\gamma > 1/(1+\epsilon_h)$,
or, equivalently, $\bar{y}_m > 0$.  The integration over $\hat{y}_{13}$
defined in Eq.~(\ref{e10}) can be rewritten as
\begin{equation}
\int d\hat{y}_{13}\,
= \int_0^1 d\hat{y}_{13}
- \int_{\bar{y}_m}^{\bar{y}_c} d\hat{y}_{13}
- \int_{1-\bar{y}_c}^{1-\bar{y}_m} d\hat{y}_{13}\ .
\label{e16}
\end{equation}
When $x_\gamma > 1/(1+\epsilon_h)$, the subtraction term on the right
of Eq.~(\ref{e3}) no longer vanishes.  Combining the real contribution
from the \underline{first} term in Eq.~(\ref{e16}), the virtual
contribution, and the subtraction term in Eq.~(\ref{e3}), we obtain
the complete partonic inclusive cross section.\cite{BGQ1}  The
\underline{third} term in Eq.~(\ref{e16}) yields contributions that
vanish when $\delta^2\rightarrow 0$.  Again, the \underline{second}
term generates both infrared and collinear sensitivity.  (The
reason that a finite contribution is present when
$x_\gamma > 1/(1+\epsilon_h)$ is that there is a mismatch between the
cone definition at the level of Feynman diagrams and
collinear fragmentation through parton-to-photon fragmentation
functions.)  Integration yields results such as
$(1/\epsilon)\delta(1-x_1)$ from $1/(1-x_1)^{1+\epsilon}$ terms, and
$\ell n(\bar{y}_m)$ from $1/y_{13}$ in Eq.~(\ref{e8}).
The $\ell n(\bar{y}_m)$ contribution is
logarithmically divergent when $x_\gamma\rightarrow 1/(1+\epsilon_h)$.

Because uncanceled infrared poles are present in the partonic
hard parts, $\hat{\sigma}_{e^+e^-\rightarrow q X}^{(1)iso}$,
we conclude that the conventional factorization theorem
for the cross section of isolated photons in $e^+e^-$ annihilation
breaks down when $x_\gamma\sim 1/(1+\epsilon_h)$. The value of
$\epsilon_h$ is chosen in individual experiments.  We turn now to a
discussion of the implications and physical consequences.

First, we caution that the breakdown of perturbative
factorization for the cross section of a physical process does not
render the process useless.  Factorization of the cross section for
massive lepton-pair production, the Drell-Yan process, fails at order
$1/Q^4$ \cite{Taylor,QS}.  Breakdown means that one
may not factorize contributions proportional to $1/Q^4$ into
infrared-safe short-distance hard parts times well-defined
long-distance matrix elements.  The terms proportional to $1/Q^4$ are
still finite, and the Drell-Yan process remains a fine process
for tests of QCD dynamics, as long as $Q^2$ is large enough.

In isolated photon production, when $x_\gamma \sim 1/(1+\epsilon_h)$,
the breakdown of factorization demonstrated in this paper means that
the cross section cannot be factored into a sum of terms each having
the form of an infrared-safe partonic hard part times a corresponding
parton-to-photon fragmentation function.  Nevertheless, the measured
cross section is still well-behaved and finite.  It becomes a task to
show that one can still extract meaningful information from the data
on isolated photons, even if factorization does not hold throughout
phase space.

In $e^+e^-$ annihilation, as long as $x_\gamma$ is kept less than
$1/(1+\epsilon_h)$, the subtraction term in Eq.~(\ref{e3}) vanishes.
The short-distance hard part $\hat{\sigma}^{(1)iso}_{e^+e^-\rightarrow
q X}$ defined in Eq.~(\ref{e3}) is then given by the isolated partonic
cross section in Eq.~(\ref{e15}).  Because
$x_1<1$ when $x_\gamma <1/(1+\epsilon_h)$, the $\delta(1-x_1)$ terms
do not contribute, and the ``+'' prescription is not relevant.  The
short-distance hard part, $\hat{\sigma}^{(1)iso}_{e^+e^-\rightarrow
q X}$, is then finite.  After the convolution over $z$
in Eq.~(\ref{e1}), terms such as $1/(1-x_1)$ and $\ell
n(1-x_1)/(1-x_1)$ produce contributions proportional to $\ell
n(1/x_\gamma-(1+\epsilon_h))$.  These are logarithmically divergent
when $x_\gamma\rightarrow 1/(1+\epsilon_h)$, but if
$x_\gamma$ is kept much smaller than $1/(1+\epsilon_h)$, the
cross section for isolated photons in $e^+e^-$ annihilation is
well-behaved.

The opportunity to extract good information on photon fragmentation
functions is one of the reasons for the study of isolated photons in
$e^+e^-$ annihilation. \cite{ISOexp,ISOthy}  Equation~(\ref{e1})
indicates that a large value of $\epsilon_h$ allows a large range of
values of $z$.  On the other hand, a large value of $\epsilon_h$
leaves a small range of $x_\gamma$ in which a fixed-order analytical
calculation can be trusted.  Therefore, special care must be taken
when data from $e^+e^-$ annihilation are used to extract photon
fragmentation functions.

For production of isolated photons at hadron-hadron colliders,
the physical cross section is obtained after an integration over
the momentum fractions of incoming partons.  One is not free
to impose a selection on $x_\gamma$ analogous to that in
$e^+e^-$ annihilation, and the integration is done throughout the part
of phase space where the breakdown of factorization discussed in this
Letter takes place.  It is therefore not altogether straightforward to
specify the precise form and magnitude of the fragmentation
contribution to isolated prompt photon production in hadron-hadron
collisions.  More discussion of this question will be found in
Ref.~\cite{BGQ2}.

\section*{Acknowledgement}

We thank George Sterman for helpful discussions.
Work in the High Energy Physics Division at Argonne National
Laboratory is supported by the U.S. Department of Energy, Division of
High Energy Physics, Contract W-31-109-ENG-38.  The work at Iowa State
University is supported in part by the U.S. Department of Energy
under Grant Nos. DE-FG02-87ER40731 and DE-FG02-92ER40730.


\vskip 0.6in

\centerline{\large FIGURE}

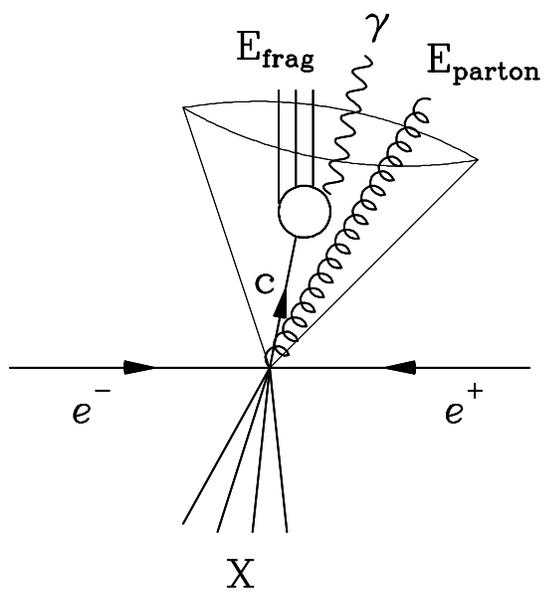
\noindent Fig.~1.\ Illustration of an isolation cone containing a
parton $c$ that fragments into a $\gamma$ plus hadronic energy
$E_{frag}$.  In addition, a gluon enters the cone and fragments giving
hadronic energy $E_{parton}$.


\end{document}